# MIS: A Multi-Identifier Management and Resolution System in the Metaverse

Han Wang[1], Hui Li[1, 2, 4, *], Abla Smahi[1], Feng Zhao[1], Yao Yao[1], Ching Chuen Chan[3], Shiyu Wang[6], Wenyuan Yang[5, *], Shuo-Yen Robert Li[6]

[1] School of Electronic and Computer Engineering, Peking University, Shenzhen, 518055, China
[2] School of Mathematics and Big Data, Foshan University, Foshan, 528000, China
[3] Department of Electrical and Electronic Engineering, The University of Hong Kong, Hong Kong, 999077, China
[4] Foshan Saisichan Technology Co. Ltd, Foshan, 528000, China
[5] School of Cyber Science and Technology, Sun Yat-sen University, Shenzhen, 518055, China
[6] University of Electronic Science & Technology of China, Chengdu, 611731, China
* Correspondence: lih64@pkusz.edu.cn; yangwy56@mail.sysu.edu.cn

The metaverse gradually evolves into a virtual world containing a series of interconnected sub-metaverses. Diverse digital resources, including identities, contents, services, and supporting data, are key components of the sub-metaverse. Therefore, a Domain Name System (DNS)-like system is necessary for efficient management and resolution. However, the legacy DNS was designed with security vulnerabilities and trust risks due to centralized issues. Blockchain is used to mitigate these concerns due to its decentralized features. Additionally, it supports identity management as a default feature, making it a natural fit for the metaverse. While there are several DNS alternatives based on the blockchain, they either manage only a single type of identifiers or isolate identities from other sorts of identifiers, making it difficult for sub-metaverses to coexist and connect with each other. This paper proposes a Multi-Identifier management and resolution System (MIS) in the metaverse, supporting the registration, resolution, and inter-translation functions. The basic MIS is portrayed as a four-tier architecture on a consortium blockchain due to its manageability, enhanced security, and efficiency properties. On-chain data is lightweight and compressed to save on storage while accelerating reading and writing operations. The resource data is encrypted based on the attributes of the sub-metaverse in the storage tier for privacy protection and access control. For users with decentralization priorities, a modification named EMIS is built on top of Ethereum. Finally, MIS is implemented on two testbeds and is available online as the open-source system. The first testbed consists of 4 physical servers located in the UK and Malaysia while the second is made up of 200 virtual machines (VMs) spread over 26 countries across all 5 continents on Google Cloud. Experiments indicate that MIS provides efficient reading and writing performance than the legacy DNS and other public blockchain-based workarounds including EMIS and Ethereum Name Service (ENS).

CCS CONCEPTS • **Networks~Network services** • Information systems~Information storage systems~Storage architectures~Distributed storage • Security and privacy~Formal methods and theory of security~Trust frameworks • Security and privacy~Human and societal aspects of security and privacy • **Social and professional topics~Professional topics~Management of computing and information systems~System management~Centralization / decentralization**

**Additional Keywords and Phrases:** Metaverse, Blockchain, Domain Name System, Identifier Management

## 1 Introduction

Neil Stephenson proposed the concept of *metaverse* in his 1992 science fiction novel named *Snow Crash* [1]. Since its inception, the term metaverse has been described in a variety of ways, such as a second life [2], 3D virtual worlds [3], and life-logging [4]. The metaverse is now often seen as completely immersive, hyper spatiotemporal, and self-sustaining virtual shared world. *Users* of the physical world access the metaverse as *avatars* in the virtual world. Users can generate content, or mint *Non-Fungible Tokens (NFTs)*, interact with other avatars and link up with digital twins of real objects, virtual objects, applications, and other accessible things. The virtual world is composed of a series of interconnected *sub-metaverses* [5]. The avatar of the user can access a variety of applications from each sub-metaverse, like gaming, social dating, virtual museums, and concerts.

The metaverse is intended to connect everything in the world, including the digital twins of physical entities and systems, the avatars of users, as well as the vast amounts of data. Therefore, the resources in the metaverse include various identities, contents, services and the data that goes with them, making up the key components of the virtual metaverse. We believe that in order to improve the user experience, a *Domain Name System (DNS)*-like system for managing and resolving resources plays a vital role as the infrastructure of the metaverse.

The DNS was originally widely used to solve the problem that IP addresses on the conventional TCP/IP network architectures are not human-friendly. The legacy DNS maintains the mapping between domain names and IP addresses and provides users with resolution services. DNS is a centralized system in terms of root zone administration even if DNS servers are distributed. In other words, DNS servers operate in reference to a central authority. More precisely, the recursive resolution of domain names is ultimately determined by the root zone, which is overseen by the Internet Assigned Numbers Authority (IANA) and the Internet Corporation for Assigned Names and Numbers (ICANN). The legacy DNS, on the other hand, is susceptible to Denial of Service (DDoS) attacks because of its centralized hierarchical structure. Therefore, in order to avoid centralization risks in resolving massive resources, the architecture of DNS alternative system should be decentralized, especially in the metaverse.

Blockchain, the underlying technology of Bitcoin, is a popular decentralized technology [6]. With the advent of blockchain platforms like Ethereum [7], blockchain applications have gone far beyond cryptocurrencies. In essence, a blockchain is a distributed ledger. The technology is nowadays impacting a variety of applications such as the Internet of Things (IoT), intelligent manufacturing, supply chains etc. In regard to blockchain-enabled metaverse applications, the hash-chained blocks and Merkle trees provide cryptographic assurance for *multiple types of identifiers (multi-identifiers)* of resources, such as identities, contents, services and IP addresses, in append-only ledgers, preventing the original data records from being tampered with. In addition, consensus protocols help to fairly generate and efficiently deliver ledger entries among multiple entities concurrently, thereby solving the problems related to trusted third parties. Due to the properties of immutability and decentralization in trustless distributed environments, blockchain can be a promising solution to the DNS alternatives in the metaverse.

The first blockchain-based DNS system, Namecoin [8], supports registration, update, and transfers similar to traditional DNS but on a Bitcoin fork. It was designed to be a more general name-value system rather than an alternative to legacy DNS. This was also the first Bitcoin-driven solution to square the Zooko's triangle [9], long-standing and persisting problem of creating a naming system that is simultaneously secure, decentralized, and human-friendly. On the basis of Namecoin, Blockstack [10] integrates DNS and Public-Key Infrastructure (PKI) and develops the so-called "virtualchain layer" to achieve good portability. The first actual naming system that is based directly on Bitcoin is said to be Blockstack. However, the most functional system, to date, is Ethereum Name Service (ENS) [11]. Unlike Ethereum, ENS places more of an emphasis on name resolution than identity management. Other initiatives, including DNSChain [12] and Emercoin [13], were built on other underlying technologies that were comparable to Namecoin or Blockstack. These initiatives have only enhanced social or economic aspects that are beyond the purview of this paper.

All the aforementioned DNS alternative systems are built on top of the public blockchain and offer decentralization advantages. However, many challenges arise when implementing a DNS alternative system based on a public blockchain in the metaverse. First, censorship of identifier compliance is challenging. The public blockchain has permissionless nodes, which makes it difficult to perform compliance checks on identifiers and their corresponding resource data during consensus. Similarly, illegal identifiers and data are difficult to seize. Second, small public blockchains are vulnerable to attacks. DNS alternatives based on these blockchains should expand their network scales or use technologies like virtualchain to switch to large public blockchains. However, even on Bitcoin, attackers with 25% of the computing power may be able to launch a 51% attack that poses threat to the security of the system [14]. Last but not least, the generation of blocks on public blockchains is limited, especially for those employing non-deterministic consensus such as Nakamoto consensus. It is possible to reduce the difficulty of mining in order to increase throughput and delay reduction, but at the expense of increased vulnerabilities. Although Ethereum 2.0 upgraded its consensus to Proof of Stake (PoS), its blockchain generation time did not decrease as a result of security concerns.

The consortium blockchain introduces constrained permission and authorization, weakening the public blockchain's decentralized features while assuring that core nodes are regulated and trusted. Deterministic consensus algorithms are usually used in the consortium blockchains to improve efficiency and reduce energy consumption. Therefore, for users who place a high priority on manageability, consortium blockchains are a promising approach for achieving efficiency and security.

Few works have attempted to integrate DNS with consortium blockchain. For instance, in order to obtain high credibility of domain name resolution results, DNSTSM [15] maintains DNS cache resources on a consortium blockchain. Such systems merely provide incremental improvement on the legacy DNS framework without solving its centralization problem. TD-Root [16] proposed a trustworthy decentralized DNS root management architecture based on a permissioned blockchain. Ho-Kyung Yang et al. [17] also proposed a solution to manage content identifiers in Named Data Network (NDN) environment.

On the other hand, with the development of new ecology and application platforms, the metaverse gradually evolves into a human-centric collection of sub-metaverse covering different types of resources. Therefore, identity identifiers that uniquely identify entities are the core of all identifiers. It has a tight bind with blockchain, which supports identity management by default. By establishing trusted digital identity identifiers based on cryptographic means in the metaverse, no matter how the resource data changes, it can be traced back to its associated encrypted address. That is, the only anchors for all entities in the metaverse are identity identifiers. As a result, it is feasible to manage multi-identifiers by identities in the metaverse via a blockchain-based DNS alternative system.

However, the majority of blockchain-based DNS alternative systems manage only a single type of identifiers, such as ENS; or manage each type of identifiers separately. For example, Blockstack creates a new namespace in addition to domain name services to provide PKI and identity management, which is arguably not enough. The relationship between these multi-identifiers and how to manage them cohesively within a system must thus be considered by DNS alternative systems.

In this paper, we have implemented MIS, a **M**ulti-**I**dentifier **S**ystem. The basic MIS is on a consortium blockchain that prefers manageability, and its Ethereum-based modification, called EMIS, prefers decentralization. Both versions are made publicly available (i.e., open source). MIS is part of the application layer of Multi-Identifier Network (MIN) architecture, in which the complete addressing and routing protocol stack of multi-identifiers has been developed on the network layer [18]. MIS records a global state of multi-identifiers, including identity, content, service, space, IP address, and domain names. Identity is the unique anchor for the various identifiers. Each individual and organization user (collectively referred to as *user* in this paper) and network device, only when owning an identifier, is allowed to apply for other types of identifiers. These users, for example, can apply for content identifiers that bind data to published media resources. They can also register service identifiers to provide various services. The following is a summary of our contributions.

- We are the first to propose a system that manages multiple identifiers concurrently across several sub-metaverses, incorporating the use of blockchain. Similar to DNS, MIS offers registration, update, revocation, and resolution services to metaverse applications for a single type of identifier.
- We design unified data structure for different sorts of identifiers in the metaverse by integrating identifier types in both present and future networks. Additionally, we build the identifier space that allows one or more identifier types in order to enable the unified management of multi-identifiers in various sub-metaverses.
- We present a lightweight scheme for on-chain data, so as to speed up reading and writing the append-only operation logs of multi-identifiers.

- To test the functionality and performance of MIS, we implemented the basic MIS and EMIS. By doing so, we set up two testbeds, one with 80 nodes on four physical servers in the UK and Malaysia, and the other with 200 nodes spread over 26 countries across all five continents using Google Cloud.

The remainder of this paper is organized as follows. In Section 2, we introduce related works of the blockchain-based DNS alternatives in the metaverse. Then, the identifier and identifier space in MIS are formally defined in Section 3. We give the design details of the basic MIS and EMIS separately in Section 4 and 5. Section 6 shows experiment results on two testbeds related to the performance of reading and writing identifiers. Section 7 summarizes the differences between MIS and other existing DNS alternatives. We conclude our work in Section 8.

## 2 Related Work

This section outlines the work related to building a blockchain-based DNS alternative system in the metaverse. In this paper, the term *identifier* means the identification information of various resource data in the network.

## 2.1 Resource Management in Metaverse

Resources such as identities, contents and services are generated and exchanged in the metaverse to form a dynamic virtual world. For example, users or avatars trade land parcels and equipment in the decentralized virtual world called Decentraland [19]. They can also construct their own buildings and social games. In Cryptovoxels [20], users or avatars are able to purchase land and build virtual shops and art galleries, where digital assets are displayed and traded. Sandbox [21] developed a blockchain-based game system that maintains users' ownership of digital lands and contents. In these projects, transaction details of resource data are recorded on the Ethereum blockchain to guarantee belonging identities. As a result, the reliable identity and resource data are essential to the continued existence of virtual worlds.

### 2.1.1 Management of Identity Identifiers

In the metaverse, a class of objects is mapped from the real world to the virtual world. For example, Haihan Duan et al. [22] implemented the CUHKSZ Metaverse, a virtual copy of the physical campus of the Chinese University of Hong Kong, Shenzhen. The consortium blockchain FISCO-BCOS is adopted as the underlying framework due to the low cost and regulability. It is necessary to identify the objects so that they can be discovered, addressed and accessed on the network. The other class of objects is produced natively in the virtual world, which also requires identity identifiers for management.

At present, identifiers are commonly adopted in the IoT field, including Bar Code, Quick Response (QR) Code, Radio Frequency Identification (RFID), etc. Considering the inherent characteristics of immersion and interoperability in the metaverse, identity identifiers can also combine biological, spatio-temporal and another attribute information. Aleksandar Jovanović et al. [23] designed a multi-dimension authentication scheme based on facial recognition, fingerprint, voice recognition, message, and mail. An access control system was then implemented in the metaverse platform *VoRtex* for identity management. The identity data, however, was located on a cloud server with centralization risks. Zijian Bao et al. [24] proposed a three-layer IoT security architecture based on blockchain to support functions of identity authentication, access control and privacy protection, but with long read and write time. Mohammed Amine Bouras et al. [25] further improved the efficiency of identity management in IoT systems and implemented it on the Hyperledger Fabric platform. Yongjun Song et al. [26] proposed a smart city standard model that provided identity management services based on blockchain, but this is only a rudimentary mode that lacks technical details.

### 2.1.2 Data Storage and Management

A natural problem in the metaverse is how to protect data of users and avatars. Several approaches to privacy protection and access control are frequently employed for cloud-based data storage and management. Chunpeng Ge et al. [27,28] combine proxy re-encryption [29] with Identity-Based Encryption (IBE) [30] or Attribute-Based Encryption (ABE) [31,32] to achieve secure, fine-grained and scalable data access control. They also propose

Ciphertext-Policy Attribute-Based mechanism with Keyword Search and Data Sharing (CPAB-KSDS) [33] to ensure data confidentiality for cloud computing. The scheme [34] based on the Revocable Attribute-Based Encryption [35] is further proposed to protect data integrity. However, these schemes are authorized by a single Attribute Authority (AA). For decentralization requirements, multi-authority revocable ABE schemes [36-38] are proposed, which use multiple AAs [39] instead of the single AA to authenticate users. In effect, a user in the metaverse only trusts the AAs in the sub-metaverses corresponding to its attributes, not all of them. As a result, these schemes cannot be directly used in MIS.

In order to save on-chain costs and improve scalability, the data storage of the metaverse system usually uses a combination of on-chain and off-chain solutions. Blockchain nodes only store the necessary metadata and the index of off-chain data, through which the associated data can be quickly located. Solutions for off-chain storage include centralized data centers, decentralized cloud storage and edge storage, InterPlanetary File System (IPFS) [40], etc. CryptoPunks stores the hash value of the metadata and media data on the blockchain, as well as integrated NFT pictures on a centralized server [41]. The integrity and reliability of pictures can be verified by comparing their hash values. However, the centralized server exposes this content resource, that is, the NFT pictures, to security threats such as loss and DDoS attacks. In Decentraland, ownership and other tradable information of NFTs are written on the Ethereum blockchain. At the same time, the user location and scenario status for real-time interaction are stored on off-chain end devices or non-core edge servers [42]. Similarly, in Sandbox, the transaction data of digital assets is stored on the Ethereum blockchain, while the associated media data and uncasted digital assets are on IPFS and Amazon's S3 cloud servers, respectively [22].

## 2.2 Blockchain-based DNS Alternatives

Blockchains are divided into public, private and consortium blockchains. This categorization is based on technical characteristics of complete decentralization, centralization and partial decentralization/multi-decentralization, respectively. Current research on blockchain-based DNS alternative systems is mainly considering public and consortium blockchains.

### 2.2.1 Solutions Based on Public Blockchains

In DNS alternative systems, Bitcoin is now the most popular underlying public blockchain. On a fork of Bitcoin, Namecoin [8] enables name-value storage and resolution. Therefore, a name can only be 64 characters long at most in order to prevent the blockchain from growing rapidly. Instead of utilizing Namecoin, Blockstack [10] implements a virtualchain on top of the Bitcoin main chain to expand storage capacity and improve security. D3NS [43] integrates a distributed hash table and domain name ownership implementation based on Bitcoin. All of the aforementioned DNS alternative systems adopted Nakamoto consensus and hence, they inherited a slow and expensive registration process. HandShake [44] adopts Bitcoin-NG [45] to improve throughput.

Since adding new features to Bitcoin is challenging, Ethereum has been introduced as a programmable and flexible solution that can extend new functions directly via smart contracts. ENS [11] is an extended smart contracts-enabled on-chain DNS mapping human-readable names to machine-readable Ethereum addresses. Stacks 2.0 [46] also incorporates general-purpose smart contracts into its original Blockstack, to overcome some of Bitcoin's limitations. Additionally, BlockZone's [47] improved Practical Byzantine Fault Tolerance (PBFT) [48] consensus mechanism provides more efficient name resolution.

### 2.2.2 Solutions Based on Consortium Blockchains

Researchers are motivated to work on the cache security of the legacy DNS since the core nodes of the consortium blockchain are regarded as reliable. DNSTSM [15] implements a secure DNS cache to prevent cache poisoning attacks. It also introduces identity-based access control to prevent potential threats from unknown nodes. Likewise, BlockNDN [49], DecDNS [50], and BlockONS [51] suggested generating a unique hash of the original data and storing it on the blockchain to avert cache poisoning. TD-Root [16] introduced a trust value and penalty mechanism to eliminate the risks of cache poisoning and trust risks.

A small number of studies concentrated on decentralizing the domain name servers leveraging consortium blockchain in addition to incrementally optimizing the classic DNS cache. Yantao Shen et al. [52] provide DNS service on a permissioned blockchain by building a TLDChain to save computing power. ConsortiumDNS [53] introduces a domain name service based on a hierarchical consortium blockchain. To alleviate storage limitations of blockchain, a 3-tier architecture was designed to separate the data and operation of domain name transactions. However, ConsortiumDNS showed a very low throughput urging the need for a consensus algorithm with less data and higher efficiency, such as Hotstuff [54] or Parallel Proof of Vote (PPoV) [55].

## 3 Formal Definitions

Before designing the architecture of MIS, it is important to first define the unified form of MIS identifiers. Then, the formal definition of the identifier space is given.

According to current metaverse applications, the main resources are identity, content, service, space, and their associated data. We analyze the identifier types in traditional and future networks. Just as IPv4 address works at the network layer of traditional network architectures, identity, content, service, and space identifiers are meaningful in future networks. In addition, the domain name is also considered as an identifier type in MIS. We define the MIS identifier and identifier space as follows.

**Definition 1: (Identifier).** *The identifier $i_j = type_j: identifier\_name$ of a user or device is defined as a combination of two string, where $type_j (j = 0,1,2, \ldots, k)$ represents the type of identifiers. The actual identifier is written in $identifier\_name$.*

**Definition 2: (Identifier Type Set).** $I = \{i_0, i_1, i_2, \ldots\}$ *represents the set of all the identifier types in the network. Where, $i_0$ is the identity of users or devices, which is the most basic and indispensable identifier. $\{i_1, i_2, \ldots\}$ optionally refer to multi-identifiers such as content $i_1$, service $i_2$, geographic location $i_3$, and IPv4 address $i_5$. In particular, $I_k$ is a subset of $I$ and if we take $k = 2$ as an example, then $I_2 = \{i_0, i_1\}$.*

**Definition 3: (Node Set).** $V$ *is the set of all nodes in the network, including end devices, hosts, routers, and switches.*

**Definition 4: (Identifier Space).** $C^{I_k} = (I_k, V_{I_k})$ *is a 2-tuple. $I_k$ represents the identifier type set in $C^{I_k}$, and $V_{I_k}$ is a subset of node set $V$. $C^{I_k}$ constitutes an identifier space when the following conditions are met:*

*(1) The identifiers and nodes in $C^{I_k}$ are defined in the network. That is, $I_k \subseteq I$ and $V_{I_k} \subseteq V$.*

*(2) The identifier types set in $C^{I_k}$ must contain the identity. That is, $i_0 \in I_k$.*

*(3) All nodes in $C^{I_k}$ own all the identifier types it contains. That is, $\forall v \in V_{I_k}$ and $i_j \in I_k$, $v$ owns $i_j$.*

*(4) Any node owning all the identifier types in the set $I_k$ belongs to $C^{I_k}$. That is, if $\exists v \in V$ and $\forall i_j \in I_k$, $v$ owns $i_j$, then $v \in V_{I_k}$.*

Figure 1 illustrates graphically how the identifier space is divided in the network. Each circle indicates the node and its identifier types. Identity ($i_0$) is the basic identifier that all nodes own, so the entire network is the identity identifier space.

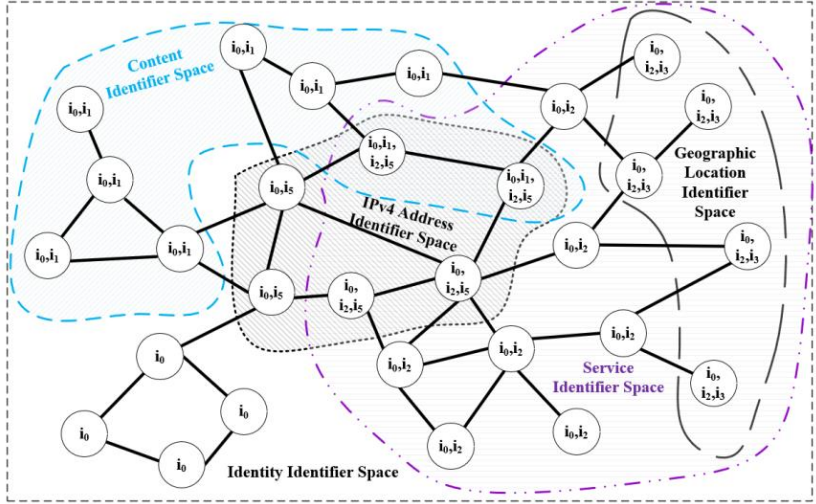
Figure 1: A schematic diagram of identifier space division.

## 4 Design of Basic MIS

For the coexistence of diverse sub-metaverses, the basic MIS achieves the simultaneous management of identifiers and their associated data from different sources on a consortium blockchain. This section describes the 4-tier architecture of MIS in detail. The managing and resolving functions of multi-identifiers are also implemented.

### 4.1 MIS Tiers

The basic MIS is based on a consortium blockchain designed to essentially maintain a global state of multi-identifiers and associated resource data among trusted core nodes. As depicted in Figure 2, it has a 4-tier architecture. Candidate nodes compete for the privilege related to the block generation process to become core nodes. The lightweight consensus algorithm is run among core nodes, logging state changes of resources in the form of transactions. Meanwhile, the metadata and complete data of resources are indexed hierarchically and eventually distributed off-chain. Such architecture provides good extensibility, and a loosely-coupled relationship between the different tiers. Therefore, it would be feasible to modify a single tier without changing the operational logic of the other tiers.

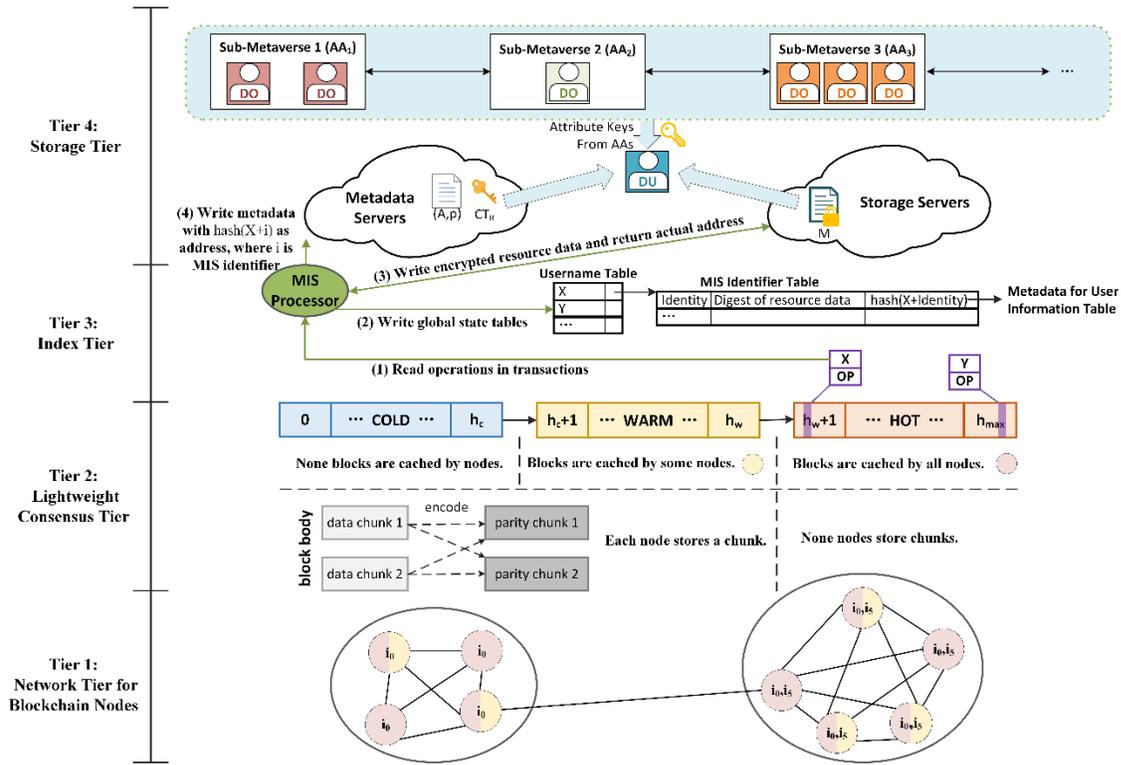
Figure 2: MIS's basic 4-tier architecture.

**4.1.1 Tier 1: Network Tier for Blockchain Nodes**

Consortium blockchain nodes occupy the lowest tier. Nodes responsible for managing the identifiers at the same level constitute the core network. There are three rights in it, which require different processing of messages and can be permuted and combined arbitrarily. That is, a core node has one or more rights.

*(1) Bookkeeping:* The bookkeeper node has the right to write transactions. The bookkeeper receives the transaction proposals and saves them into its local pool. After starting consensus round, the bookkeeper selects part of the transactions as a set $Tx\_List = \{t_1, t_2, \cdots, t_{total}\}$ and, together with the block height $h$, timestamp $Ts$, and the hash of the previous block $Pre\_Hash$, it generates the signed message $\langle Prepare - Tx\_List \rangle$. The prepare message $\langle Prepare - Tx\_List \rangle$ is then broadcasted to the consortium blockchain network for collective voting. If approved, the set of transactions $Tx\_List$ will be appended in the highest block.

A term-of-office system governs bookkeeping rights. At the end of the term, voters rate the workload of the bookkeepers and submit the scores to the aggregator. By counting the total scores, the aggregator ensures that the top $ratio_{bookkeeper}(\%)$ of the scores are consistently re-elected, while the bottom $(100\% - ratio_{bookkeeper})$ are replaced by the oldest nodes in the candidate queue to balance the activity and stability of nodes. The list of new bookkeepers is written into a special transaction $Tx\_Permission\ (Bookkeeping)$ to commit the transition of bookkeeping rights.

*(2) Voting:* The voter node has the right to vote. The voter votes on the transaction sets $Tx\_Lists$ from bookkeepers in the consensus round. If some bookkeepers are malicious or crashed, the voter cannot collect all the sets. In this case, it needs to wait for a threshold time before voting partially on the received sets. Specifically, the voter checks them one by one, voting 1 for passed, $-1$ for failed, and 0 for unreceived. Further, the set of votes $Vote\_List$ of all the received transaction sets is compressed into a long integer and signed, and is then sent to the aggregator together with the digests $TX\_Hash$ as the vote message $\langle Vote \rangle$.

A node applies to become a voter by a special transaction $Tx\_Permission\,(Voting)$. Unlike normal transactions or $Tx\_Permission\,(Bookkeeping)$, the transaction $Tx\_Permission\,(Voting)$ requires the approval of all the voters before it can be appended on-chain. If the transaction is committed, the node will have the right to vote starting from the next consensus round, and the number of voters will increase by one. This procedure is similar to the Byzantine Fault Tolerance (BFT) consensus, but with the strictest voting threshold of 100%. This translates into a positive view of voters.

*(3) Aggregating:* The aggregator node has the right to count votes. The aggregator collects and aggregates long integer $Vote\_Lists$ from voters. Considering that both bookkeepers and voters are Byzantine, the aggregator similarly has a finite wait. As soon as the threshold time is reached, the aggregator will start to count the received votes as $Vote\_result$. The transaction set that gets approval or disapproval from more than 2/3 of all voters is considered to be determinably valid or invalid. Otherwise, the aggregator will request missing votes. In addition, since all the original votes are signed by voters, the aggregator compress the votes and signatures (Section 4.1.2). The commit message $\langle Commit \rangle$, including $Vote\_result$, compressed votes and aggregate signature, is then generated and broadcasted across the network.

Aggregation is the most important right of nodes in the network. Because only one node acts as the aggregator in a consensus round, the system services will be seriously affected if the aggregator fails. Therefore, malicious nodes should be prevented from having this privilege as much as possible. In our current implementation, the aggregator is periodically chosen among honest voters. Specifically, we select the next aggregator number $Agg$ by XOR and modulo operations based on the last vote $Vote\_List_{max}$ and its receipt timestamp $Vote\_Ts_{max}$ of the highest block, that is, $Agg = StrToInt\left(SubStringEnd32\left(hash(Vote\_Ts_{max} \oplus Vote\_List_{max})\right)\right)\ mod\ N_{voter}$. It is unlikely that an attacked voter will be selected as the next aggregator $Agg$, since the receipt time of votes is unpredictable and $Agg$ is randomly selected. In particular, once an aggregator is found to be malicious, it will be penalized and all its rights will be revoked.

*(4) No right:* The ordinary node in the consortium blockchain network without rights do not participate in the consensus process. This node has however the ability to request or receive the transaction set and the block header in a way similar to the bookkeeper and the voter. When receiving a transaction set $Tx\_List$ from a bookkeeper, nodes need to check the identifiers in it, as well as the bookkeeping right and the signature of the bookkeeper. When receiving the commit message $\langle Commit \rangle$ from the aggregator, nodes should check whether all the transaction sets $Tx\_Lists$ mentioned in the block header $BH$ have been received. The missing sets will be obtained by requesting them from the neighbors. Finally, they assemble the complete block, and store it based on the timeline strategy (Section 4.1.2).

**4.1.2 Tier 2: Lightweight Consensus Tier**
The lightweight consensus tier is located above the network tier for consortium blockchain nodes. The basic MIS uses PPoV [55], an efficient BFT consensus algorithm, for writing and verifying operation transactions such as registration, update, revocation, renewal, and transfer to the blockchain. The consensus process is described in Section 4.1.1. The advantage of using PPoV in a DNS alternative system is that blockchains are increasingly seen as the underlying communication channel for announcing state changes. The consensus can only serve for instrumental ordering. PPoV consensus separates a voting right to enable voters to vote for transactions in the blockchain according to their own rules. As a result, the consensus results will be more reasonable.

However, all votes and signatures are stored as proof in the block header, thus consuming a large amount of storage space. Another consideration is that each bookkeeper signs its generated transaction set. We lightweight the PPoV consensus algorithm in three aspects.

The first is, to solve the problem of large signatures, we use the Boneh-Lynn-Shacham (BLS) signature scheme [56] instead of the Elliptic Curve Digital Signature Algorithm (ECDSA) to aggregate signatures. Another optimization is to redesign the structure of the PPoV block, as shown in Figure 3, to reduce the size cost.

| **Block Header** | Height | Pre-hash | Timestamp |
|---|---|---|---|
| | Vote_result | Votes | Voters' Aggregate Signature |
| | Bookkeepers' Aggregate Signature | Merkle Root_0+Tamestamp_0 | |
| | | ... | |
| | | Merkle Root_h+Tamestamp_h | |
| **Block Body** | Tx_Lists{0,...,h} | | |

Figure 3: Redesigned lightweight block structure of PPoV.

For a single block, optimizing the signature and structure is a viable means of lightweighting. Theoretically, the more bookkeepers or voters there are, the larger the size of signatures in the original PPoV. However, the aggregate signature with constant size in the redesigned block header is insensitive to the number of nodes. As a result, the redesigned lightweight block is significantly smaller than the original PPoV.

The third aspect is to reduce redundancy. Generally speaking, each node has to store the full data, which causes storage stress for the nodes. However, the frequency of querying old blocks is low. Assume that there are $n$ nodes in the consortium blockchain, no more than $f$ of which are Byzantine ($n > 3f$). Given that data in the block header is accessed frequently by the nodes, we only compress actual transactions in the block body using $(n - 2f, 2f)$ Reed-Solomom (RS) code [57], an erasure code. It has the advantage of recovering all the original data chunks with only $(n - 2f)$ chunks, which is suitable for blockchain with Byzantine nodes.

Since transactions in newly generated blocks are in general visited frequently, compression of this part of the data leads to low query efficiency. Therefore, we propose a timeline strategy for lightweight storage. All blocks are divided by heights into hot, warm, and cold states, with the newest block being hot and the oldest being cold.

*(1) Hot blocks* are frequently accessed, so each node should cache a copy. Let us take the length of hot space as $L_h$ and the latest block height as $h_{max}$, then the blocks with the heights in the range of $(h_{max} - L_h, h_{max}]$ are in a hot state.

*(2) Warm blocks* are between hot and cold blocks, and they have a certain probability of being accessed. Let the length of warm space be $L_w$, so that blocks with heights in the range of $(h_{max} - L_h - L_w, h_{max} - L_h]$ are in a warm state. When the state of a block changes from hot to warm, each node first encodes the warm block to $(n - 2f)$ data chunks $\{C\}_{i=1}^{n-2f}$ and $2f$ parity chunks $\{C\}_{i=n-2f+1}^{n}$. Then, the block header $BH$ and one of the chunks are uniquely stored by the node. If the chunk fails, the node can request other arbitrary $(n - 2f)$ chunks to decode to the original block.

In order to reduce the probability of decoding, we assign some nodes to cache the warm blocks. Suppose a block is cached on $n_w$ nodes. The decoding process is triggered only when all the $n_w$ caches are down. We believe that the probability of a warm block being accessed decreases over time, and therefore the number of caches should decrease. The latest warm block is cached on $(2f + 1)$ nodes, because the number of failed caches is no more than $2f$, and transactions can be queried without decoding. The oldest warm block is cached only on one node to ensure that it does not take up extra storage space. The relationship between cache quantity $n_w$ and block height $h$ is that $n_w(h) = \left\lfloor (h - h_{max} + L_h + L_w - 1) \cdot \frac{2f}{L_w - 1} \right\rfloor + 1$.

*(3) Cold blocks* have a low probability of being accessed, and their heights belong to the range $[0, h_{max} - L_h - L_w]$. At this stage, nodes only hold block headers and chunks generated in the warm space and remove caches. However, given the fact that there may be some important old blocks that are frequently visited, the node monitors the number of visits to each block. For cold blocks that are too frequent, the node keeps its cache from being deleted.

For the general case, we theoretically calculate the total on-chain storage cost to demonstrate the effectiveness of the coding and timeline strategies. Let the size of the redesigned block head be $size_{BH}$. Suppose the block body is filled with transactions of size $size_{BD}$ ($size_{BD} \gg size_{BH}$). For the hot space with full caching, the storage cost is $size_h = L_h n(size_{BH} + size_{BD})$. In warm space, blocks are both encoded and cached. Among them, the size of the encoded data chunks is $size_w^{chunk} = L_w n \left(size_{BH} + \frac{size_{BD}}{n-2f}\right)$, and that of the cached copies is $size_w^{cache} = \sum_{h=h_{max}-L_h-L_w+1}^{h_{max}-L_h} n_w(h)(size_{BH} + size_{BD}) \leq \left[\sum_{h=h_{max}-L_h-L_w+1}^{h_{max}-L_h}(h - h_{max} + L_h + L_w - 1) \cdot \frac{2f}{L_w-1} + L_w\right](size_{BH} + size_{BD}) = L_w(f+1)(size_{BH} + size_{BD})$. Therefore, the cost of the warm space is $size_w = size_w^{chunk} + size_w^{cache} \leq L_w\left[(n+f+1) \cdot size_{BH} + \frac{nf+2n-2f^2-2f}{n-2f} \cdot size_{BD}\right]$. For the cold space with full encoding, the cost is $size_c = n(h_{max} - L_h - L_w + 1)\left(size_{BH} + \frac{size_{BD}}{n-2f}\right)$. In summary, the total cost is

$$size_{total} = size_h + size_w + size_c$$
$$\leq [n(h_{max}+1) + L_w(f+1)]size_{BH} + \left[\frac{n}{n-2f}(h_{max}+1) + \left(n - \frac{n}{n-2f}\right)L_h + (f+1)L_w\right]size_{BD}. \quad (1)$$

In order to compare with the original storage cost of on-chain data $size'_{total} = n(h_{max}+1)(size_{BH} + size_{BD})$, we calculate their difference as

$$\Delta_{size} = size'_{total} - size_{total} \geq \left[\left(n - \frac{n}{n-2f}\right)(h_{max} - L_h + 1) - (f+1)L_w\right]size_{BD} - (f+1)L_w size_{BH}. \quad (2)$$

Since $n > 3f$ and $size_{BD} \gg size_{BH}$, we get

$$\Delta_{size} \gg (f+1)(L_w + 3L_c)size_{BH}, \quad (3)$$

where $L_c$ is the length of cold space.

Equation (3) proves that the coding and timeline strategies greatly reduce the storage consumption of on-chain data. This is mainly due to the compression of warm and cold spaces. As the number of blocks increases, the overall compression ratio approaches that of encoded block body, which is $1/(n-2f)$.

### 4.1.3 Tier 3: Index Tier

The index layer maintains a global state for the basic MIS, including who owns which identifiers and how to access off-chain resource data associated with those identifiers.

An object in the metaverse owns multiple identifiers, which are dynamically added, updated or deleted as the network changes. We use a non-relational database that holds a username table and multiple MIS identifier tables by key-value pairs. A username record corresponds to a MIS identifier table, including the binding records of multi-identifiers (only identity is mandatory) as well as hashes for the username and identifier (as the metadata address for associated resources).

The MIS processor in this tier is also responsible for handling the management logic of identifiers (Section 4.2). Figure 2 uses the identity of a user as an example, where the user information table stores the data associated with such an identifier. The MIS processor first finds the identity registration operation of username $X$ in the blockchain. After that, it adds a username record $X$ to the username table and creates a new MIS identifier table of $X$ with identifier record $Identity$. It also stores the address of metadata as $hash(X + Identity)$ and the digest information of resource data. Then the MIS processor writes the encrypted user information table to the storage servers and the metadata into the metadata servers (Section 4.1.4), so that the data separation of identity identifier and associated data can be achieved for fast indexing and verifying.

In the basic MIS, each object must have an identity identifier. An object's identity is uniquely identified in the virtual world of metaverse so that this identity plays the role of "passport number". In addition, for other types of identifiers, identity serves as a unified base identifier and an anchor point.

### 4.1.4 Tier 4: Storage Tier

Storage is the top-most tier of the system, which stores the complete off-chain resource data associated with multi-identifiers. Resources containing personal information, such as the user information table, should be encrypted in this tier for privacy protection. On the other hand, the storage tier supports a variety of storage systems, devices, and replication strategies, so that metadata is stored and managed uniformly across distributed metadata servers by the MIS processor. Metadata files contain storage addresses, authentication information, encrypted keys, and access control policies for actual data or service resources. In addition, the metadata servers should have all types of identifiers to ensure that all users have access to them.

Access control in the metaverse should satisfy decentralization. At the same time, the user's characteristics related to sub-metaverses (such as biological and spatio-temporal) are always changing dynamically. To this end, we adopt a data storage and access scheme for multiple sub-metaverses with multi-authority ABE, which enables secure and controllable off-chain storage.

In this scheme, a Data Owner (DO) through the MIS processor encrypts the resource data with a symmetric key, which is encrypted with an access control policy. Considering the characteristics related to sub-metaverses as attributes, only Data Users (DUs) who satisfy the policy and possess the corresponding attribute keys are permitted to decrypt the resource data. To ensure that attributes are managed fairly, we set an AA for every sub-metaverse. The AA authenticates the user and generates the corresponding attribute keys. The scheme consists of four modules and is described below.

*(1) Initialization:* The MIS processor inputs the security parameter $\lambda$ and randomly generates the global parameter $GP$ locally. The AA of the sub-metaverse $m_i$ ($i = 1,2,3,...$) generates its key pair ($apk_i, ask_i$) according to the global parameter $GP$. $GP$ and $apk_i$ are then broadcast on the network.

*(2) Attribute key distribution:* The AA of the sub-metaverse $m_i$ inputs the global parameter $GP$, a DU's attribute $att_j$ ($j = 1,2,3,...$), the DU's identity identifier $X$, the time $period$, and its private key $ask_i$, then outputs the attribute key $sk_{att_j,X}^{period,i}$. The time $period$ limits the lifetime of the attribute key. The AA updates the key as soon as it expires. It is important to note that illegal DUs will be added to the revocation list $rl_i$ without access to subsequent attribute keys.

*(3) Encryption:* The MIS processor randomly selects a symmetric key $R$ to encrypt the resource data $M$. Then, it generates the access structure $(A, \rho)$, where $A$ is the access control policy matrix and $\rho$ is the mapping function between attributes and rows of the matrix. Next, the symmetric key $R$ is encrypted using the access structure $(A, \rho)$, the time $period$, and the public key $\{apk_i\}$ to obtain the ciphertext $CT_R$. The MIS processor writes $(A, \rho)$ to the metadata server together with the ciphertext $CT_R$.

*(4) Decryption:* When requesting resource data, the DU first accesses the corresponding metadata file. Then it verifies whether its attribute set $\{att_j\}$ satisfies the access structure $(A, \rho)$. If satisfied, it will use the attribute key set $\{sk_{att_j,X}^{period,i}\}$, the identity identifier $X$, the time $period$, and the global parameter $GP$, to decrypt the ciphertext $CT_R$ and get the key $R$. At the same time, the DU can receive the encrypted resource data from the address in the metadata file. For large resources, encrypted data blocks are read simultaneously to maximize efficiency. Eventually, the original plaintext of the resource data is recovered by decrypting the encrypted version with the key $R$.

## 4.2 Multi-Identifier Management and Resolution System

The basic MIS uses its 4 tiers to implement a complete DNS alternative system with multi-identifiers. Its main functions are registration, update and revocation. Resolution of a single type of identifiers and inter-translation across identification spaces are also supported. In addition, MIS is compatible with the legacy DNS.

### 4.2.1 Registration, Update, and Revocation

Identity is the anchor identifier in MIS. The identity of the device is assigned, while the user needs to register. Before registering an identity, the user first generates a pair of keys with appropriate strength. To register a public key as an identity, a user initiates an identity registration request, which includes the username, identity, Time-to-Live (TTL), registration fee, timestamp, signature, and other information listed in Table 1. It is then submitted to the bookkeeper of the identifier space. The user can select any bookkeeper for such identifier type, because the entire network is an identity space. When the bookkeeper receives the request, it checks the format and content, including but not limited to checking whether duplicate valid identifiers already exist in the global state tables. If successful, the bookkeeper will preferentially encapsulate requests with high registration fees into transactions and put them into the local pool for lightweight consensus. When the consensus is passed, the MIS processor will extract usernames and other required fields of newly valid identities from registration transactions and write the global state tables. If the *AboutMe* field is also populated, it is also extracted and stored as an off-chain user information table via encryption.

Table 1: Field in an identity registration request.

| Field Name | Data Type | Description |
| --- | --- | --- |
| Username | String | *Required. |
| Identity_Identifier | String | *Required. The format is "type0:"+"public key of user" |
| AboutMe | String | Optional, resource data. |
| Hash_Identity_Identifier | String | *Required, the address of the metadata. |
| Digest | String | Optional, digest of resource data. |
| TTL | Int | *Required, the valid period of the identity identifier. |
| Fee | Float | *Required. The registration fee can be zero. |
| Timestamp | Double | *Required, the generation time of the identity registration request. |
| Signature | String | *Required. The user signs with the private key. |

After a user successfully obtains an identity, it can register other multi-identifiers such as content and service. For the identifiers within the validity period, only the owner can be allowed to operate on it, for example, update or revoke. Expired or revoked identifiers can also be re-registered. The operation process is similar to identity registration. It is noted that all operations are performed only after achieving the consensus in the consortium blockchain.

**4.2.2 Resolution**

When the network only holds a single identifier space, basic MIS provides an identifier resolution service like DNS. When the identifier is resolved in this case, users can access the storage server or the associated resource. The specific resolution process is as follows.

Step 1: The user queries the global state tables to obtain the address of metadata $hash(Username + Identifier)$ and the digest information of the associated resource.

Step 2: According to the metadata address, the user accesses the metadata server and receives the metadata file as a response. This file includes the location of the storage servers and other metadata information such as the access control policies and the symmetric key.

Step 3: After receiving the metadata file, the user requests the encrypted resource data associated with the identifier from the network. The process of requesting resources includes push and pull methods, both of which have been integrated into Multi-Identifier Router (MIR). It is important to note that with pull transport, the user does not actually access the storage server location recorded in the metadata file.

Step 4: When the user receives a resource associated with the identifier, it decrypts it with the symmetric key and verifies the integrity based on the digest information. If verified, the identifier will be considered to be resolved. Otherwise, return to Step 3 and continue the request.

Although IP address identifiers in MIS are different from IP addresses in the traditional network, MIS is still compatible with legacy DNS. This is because the domain name is defined in MIS as an identifier type 6, which is bound to website resources. The IP addresses mapped to domain names are stored in the associated metadata file. In addition, we register the special usernames in the form of $DNS_{cache}: i$ $(i = 1,2,3, …)$ in the basic MIS to cache DNS resources.

When a user or device wants to resolve a domain name, it queries the MIS identifier table linked by the username record ($DNS\_cache: i$). If the query fails, the MIS processor will forward the query request in a typical form to the DNS server and cache the DNS response into the metadata file which is associated with the domain name identifier in the global state tables for future queries.

Given that domain names are actually controlled by DNS servers rather than users or devices in the basic MIS, the caching process does not wait for consensus, unlike the registration of identities and other identifiers. In other words, the credibility of the resolution service for domain names depends on the DNS itself.

### 4.2.3 Inter-translation

In addition to TCP/IP networks, there are many future networks with different types of identifiers (and corresponding communication modes) to satisfy new functions or requirements. For example, content identifiers have an advantage in data resources such as videos and pages. Service identifiers improve the flexibility, loose coupling and reusing properties of services. Identity and location identifiers are suitable for mobile devices that constantly switch locations.

Therefore, resource providers in the metaverse are encouraged to apply for identifiers and publish resources as much as possible to form multiple identifier spaces. Through the inter-translation service, the basic MIS supports users to resolve all types of identifiers and obtain resource data. Figure 4 illustrates the detailed process.

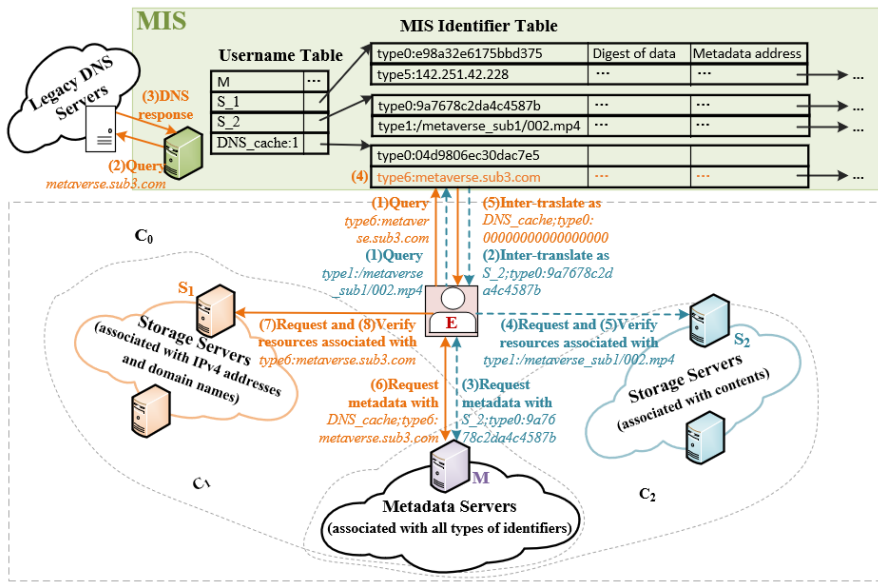

Figure 4: An example of inter-translation.

In Figure 4, we assume that the identity identifier space $C_0$ holds identity ($i_0$). In addition to the basic identity ($i_0$), $C_1$ also holds IPv4 address ($i_5$) and domain name ($i_6$). $C_2$ holds identity ($i_0$) and content ($i_1$). Metadata servers own all types of identifiers ($i_0, i_1, i_5, i_6$). A normal individual end user $E$ has only the required identity and is in the identifier space $C_0$.

In case user $E$ wants to resolve the domain name ($type6: metaverse.sub3.com$) of the identifier space $C_1$, it first queries in the global state tables. If fails, the MIS processor will perform the resolution process (Section 4.2.2) and response the user $E$ with the inter-translating result, including the username ($DNS\_cache: 1$) and its identity ($type0: 04d9806ec30dac7e5$). At this point, the domain name ($type6: metaverse.sub3.com$) has a form that can be recognized in the identifier space $C_0$. Then, the user $E$ requests the metadata server $M$ for the IPv4 address ($type5: 142.251.42.228$) in the metadata file. Finally, it communicates to the storage server $S_1$ with the identity ($type0: 04d9806ec30dac7e5$) and IPv4 address ($type5: 142.251.42.228$) for routing, so as to obtain the associated resource for the domain name ($type6: metaverse.sub3.com$).

Likewise, user $E$ can resolve the content ($type1:/metaverse\_sub1/002.mp4$) of the identifier space $C_2$. The difference between this process and IPv4 address of $C_1$ is that the data of identity and IPv4 address is obtained in push mode, while the data of content is obtained in pull mode. Therefore, the resolution process involves changing the communication mode. MIN has realized a communication mode that supports both push and pull semantics in the protocol stack. Simply speaking, one or more edge routers are set up in each identifier space, which are responsible for processing packets sent to other identifier spaces, including recognizing identifier types and changing semantics.

In conclusion, the inter-translation service enables the existence of the same resource data in multiple forms (identifiers) and the accessibility for users of different identifier spaces of sub-metaverses.

## 5 Migrating MIS to Ethereum

The MIS described in Section 4 is based on the consortium blockchain. However, most metaverse projects utilize Ethereum due to its outstanding features. Therefore, in order to accommodate the Ethereum environment, we modified MIS architecture due to loose couplings, and proposed Ethereum MIS (EMIS). The lower two tiers of EMIS follow Ethereum, including node classification and PoS consensus. The storage tier of the basic MIS is adopted at the top.

The intermediate index tier is responsible for core logic related to the management of identifiers. We redesign this tier in EMIS with Solidity smart contracts, consisting mainly of an EMIS contract and identifier contracts, as shown in Figure 5. Below is a description of the two types of contracts.

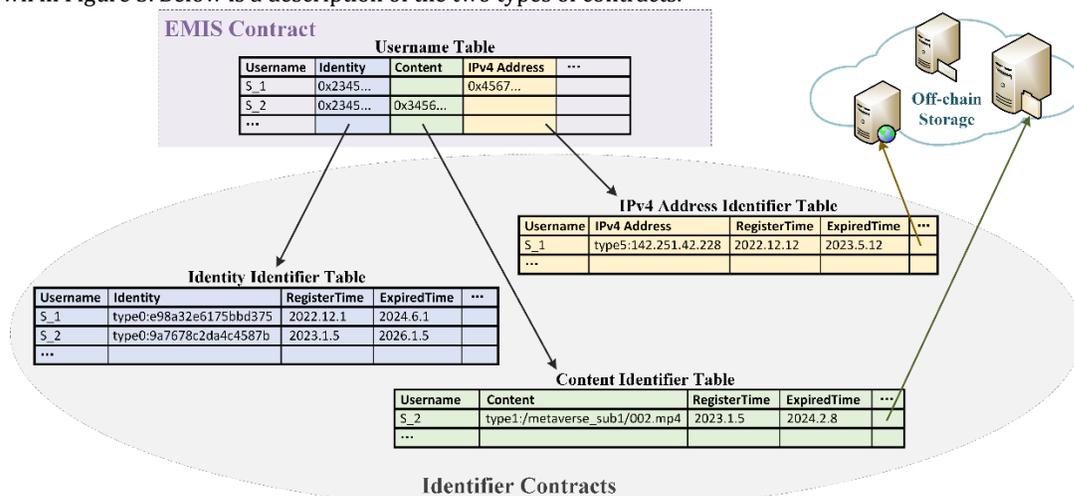

Figure 5: Solidity smart contracts in EMIS.

## 5.1 EMIS Contract

As with the basic MIS, which uses the public key as identity identifier, EMIS uses its identity to identify a specific Ethereum address. It is necessary to use usernames in order to simplify the identification of identities. Specifically, the EMIS contract maintains the username table, as well as supporting registration, update, revocation, transfer and authorization of usernames. It also integrates the interfaces of multiple identifier contracts.

A user wishing to use an identifier service must have a username. Upon registering a username, the EMIS contract filters out the sensitive words to ensure compliance without disclosing inappropriate information. After submitting the username information, the registration process will be completed once the block has been confirmed. In particular, in order to access certain sub-metaverse resource data, a user must associate their username with relevant attributes. The owner of a username will then have the right to update and revoke the username at any time. A username may also be transferred to another user, which means that its ownership is completely changed. EMIS provides authorization support, which is the most significant addition compared to basic MIS. Alternatively, a user may delegate the management of their username to a third party. Under the control of the EMIS contract, the username owner can terminate the delegation and reauthorize a new user at any time.

Another added feature is to prevent squatting. This is in view of the fact that the public blockchain is an open network and even functioning nodes may be motivated by self-interests. In this way, squatting is easily accomplished without the need to pay for the processing of a username. We refer to the two-phase mode of ENS as a means for alleviating the squatting problem. To generate commitment, a hash of the pending username, user address, and secret is created. At this point, the transaction records only the *commitment* and does not reveal any other information. In the second registration phase, the transaction records the actual pending username and secret. The *commitment'* is then computed and compared to the *commitment*. Only the matched user will be allowed to successfully register the username.

## 5.2 Identifier Contracts

A resource in EMIS is uniquely associated with an identifier. To manage and resolve identifiers of different types, we design corresponding contracts, such as the identity identifier contract, the content identifier contract, and the service identifier contract. Each identifier contract maintains an identifier table. A number of specific functions are available, including registration, update, revocation, renewal, transfer and resolution of identifiers. Moreover, the interfaces of the identifier contracts are integrated into the EMIS contract in order to make it easier for the user to call functions uniformly.

When a user with an identity wants to publish a resource, it has to register an identifier of the corresponding type. Optionally, the identifier can be generated automatically. After submitting the resource information, the user is required to pay for the associated identifier. The fee consists of two components: a processing and a Gas fee. The former is to compute the validity period of the identifier, and the latter is to execute the identifier contract.

Similar to basic MIS, an identifier can be updated or revoked during its validity period. Consequently, it is not possible to modify an expired identifier. However, a user can extend the TTL of its living identifier with a renewal service. Note that the operator is either the owner or the authorized user.

A user can also resolve a living identifier. An identifier of the form defined in Section 3 consists of two strings, the type and the actual identifier. The process of resolving an identifier involves three steps. First, the EMIS contract selects the corresponding identifier contract based on its type. The identifier contract then obtains resource information from the identifier table based on the actual identifier. Third, the user checks its attributes and accesses the metadata and resource data associated with the identifier off-chain. The original plaintext of the resource data is only available to users whose attributes match.

## 6 Experiment

In this section, we evaluate the performance of MIS to demonstrate that it can operate at competitive rates. We implement the basic MIS in Golang [58] and EMIS in Solidity [59]. We also set up a testbed for MIS on 4 physical servers located in the UK and Malaysia. Each server has two 8-core CPUs and 10 Gbps network bandwidth, and contains 20 nodes. For basic MIS, a node is simulated as a representative of a consortium country or top-level organization, with bookkeeping and voting rights. The top 90% of the scores are consistently re-elected nodes, and the bottom 10% are replaced periodically. Each node holds aggregation permission in turn. We allow no more than 100,000 transactions to be encapsulated in a block. All transactions and blocks are signed by bookkeepers using the BLS signature scheme. Each transaction and block header consists of 40 bytes and 229 bytes, respectively. Moreover, both hot and warm spaces have a length of 100. For EMIS, a node is simulated as a *Geth* client on the Ethereum network. For comparison, we deploy both EMIS and ENS, the best-known DNS alternative on Ethereum. The read and write performance of metadata and resource data depends on the off-chain storage system, which becomes a deterministic load for MIS.

**Read time:** Read time is related to the data structure and indexing method. In Figure 6, we measure the average response time for multi-identifier resolution requests. For 20,000 registered identifiers, the average time required to query one in MIS is only 34.035 milliseconds, and this meets the actual query performance of the legacy DNS (48 milliseconds).

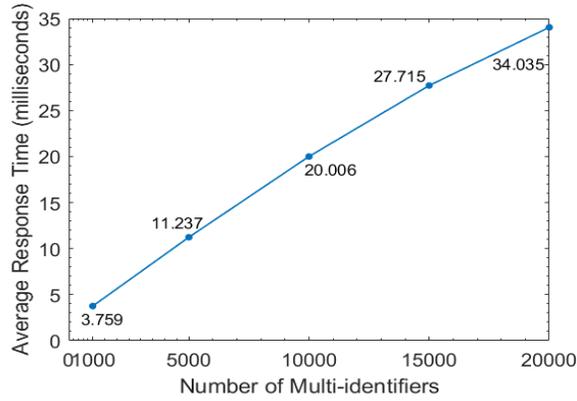

Figure 6: Response time for resolving multi-identifiers.

**Storage space:** For the basic MIS, we test the total cost of on-chain data as shown in Figure 7. With conventional PPoV consensus, each appended block adds 320.02 Megabytes to the storage cost. In the lightweight scheme, the warm and cold spaces reduce it to 36.71% (117.49 Megabytes) and 3.61% (11.56 Megabytes), respectively. As the blockchain continues to operate, the number of the most compressed cold blocks gradually increases. Therefore, with increasing block height, the compression effect of on-chain data in lightweight consensus becomes more pronounced.

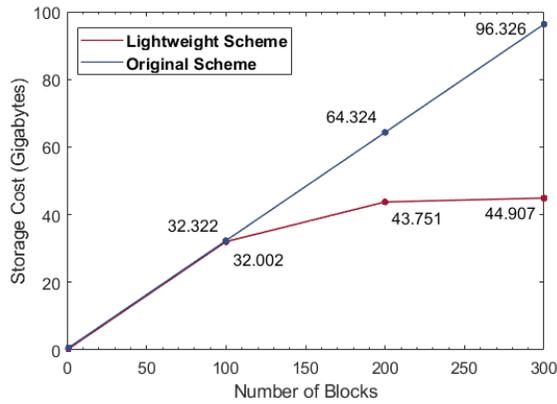

Figure 7: Total storage cost of on-chain data for different numbers of blocks.

**Write time:** In our experimental blockchain for basic MIS, at an average rate of 1,000 per second, users randomly send writing operation requests for registration, update, and revocation of multi-identifiers to the consortium nodes. The time spent in each phase of writing operations is shown in Figure 8. We obtain a total average latency of 356.90 milliseconds, significantly lower than the consensus latency of EMIS and ENS, which is an average of 13.38 seconds. The spikes at block heights 146, 169, 303 and 361 are caused by unstable bandwidth speeds between servers in different regions. In addition, because aggregators can be malicious through inaction, forgery of votes, or statistical errors, we disable the duty aggregator at block height 459 to simulate its deliberate inaction. The consensus process is halted for a period of time (a threshold), here set at 1 second. Following the timeout, the next aggregator is selected to continue this consensus round. The imperfect synchronization of the network, however, does not allow all nodes to update the aggregator immediately. Therefore, the subsequent consensus remains affected until block height 468.

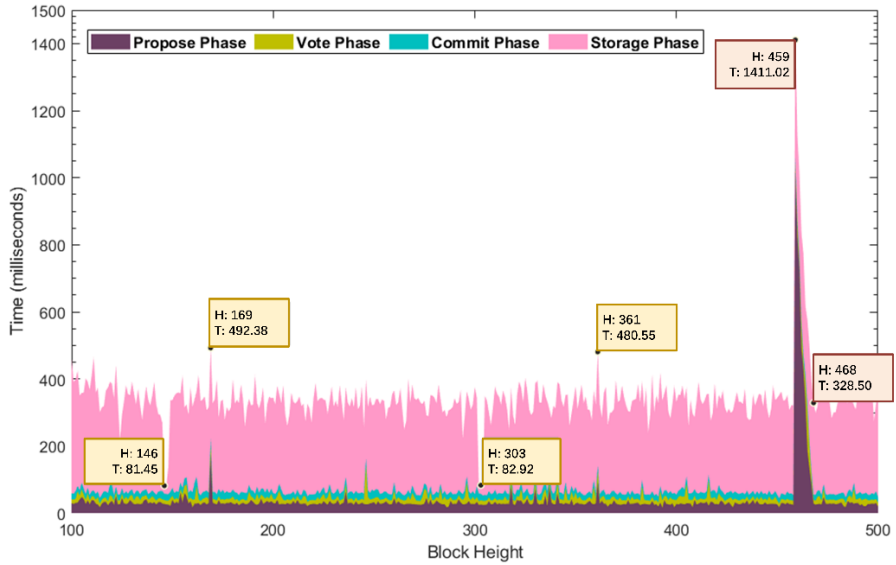
Figure 8: Write time of the basic MIS on the 80-node testbed in the UK and Malaysia.

In Figure 8, the main time cost is spent on storing blocks into the disc (presented as Storage Phase), in an average of 284.58 milliseconds. This is limited by the speed at which the CPU reads and writes memory. The consensus phase, including propose, vote and commit phases, takes only 73.32 milliseconds, accounting for 20.54% of the total latency. This is because the communication bandwidth between nodes within the server was considered as infinite on the testbed. However, performance will be largely network-bound in practice.

To more realistically verify the writing performance of the basic MIS, we simulate a globally distributed deployment on the Google Cloud. There are 200 VMs located in 26 countries from 5 continents in our simulation environment, guaranteeing that all countries can participate in the management of multiple identifier spaces. 6 to 9 VMs are deployed in each country as TLD consortium blockchain nodes for basic MIS. In total, there are 16 VMs in North America, 64 in South America, 58 in Europe, 46 in Asia and 16 in Oceania. Each VM is configured with 2 vCPUs and 4GB memory. In addition, we use speedtest [60] to measure the bandwidths of VMs in each country. The average uplink bandwidth is 897.84Mbit/s and the downlink is 1525.60Mbit/s. Other parameters are the same as the above 80-node testbed.

We evaluate the average latency in each phase of the writing operations by gradually increasing the number of nodes from 50 to 200, as shown in Figure 9. Although it takes a little longer in the consensus phase as the number of consortium nodes increases, the consensus time is still acceptable. This can be referred to the deterministic consensus used in the basic MIS. For example, the original Blockstack overlays Bitcoin with a latency of 600 seconds, whereas the basic MIS latency is even less than 10 seconds.

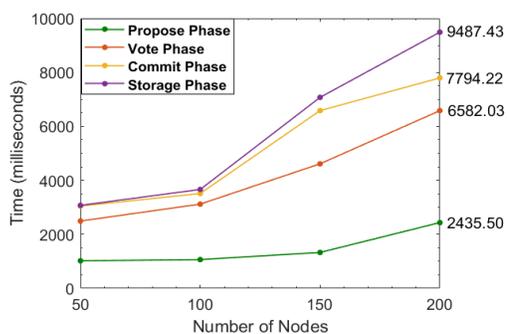
Figure 9: Write time of the basic MIS on the testbed located in 26 countries from 5 continents.

**Gas cost in Ethereum:** Gas is needed to send and execute transactions on the Ethereum blockchain. The more complex the contract, the more Gas it consumes. Table 2 compares the Gas cost of EMIS and ENS for registration and update operation requests. It is shown that EMIS is less complex than ENS with different data structures. EMIS transactions are more likely to be packed into blocks than ENS transactions with the same Gas price, as miners prefer to profitably exploit the target Gas size of a block. In addition, EMIS also supports the revocation with 60,338 Gas.

Table 2: Gas cost for writing operation requests in EMIS and ENS.

| Identifier Service | | Registration | Update | Revocation |
|---|---|---|---|---|
| Gas | EMIS | 43901 | 49875 | 60338 |
| | ENS | 56275 | 56597 | / |

# 7 Comparative Discussion

In order to acquire decentralized DNS alternatives, public and consortium blockchains are two feasible means. Bitcoin is the largest, most secure, and most actively maintained blockchain among the most suitable public blockchains for carrying decentralized applications and services. However, public blockchains including Bitcoin still pose several challenges when used for managing identifiers, as described below.

*(1) Challenge 1: Compromise Between Efficiency and Reliability*

The consensus algorithms of public blockchains face a serious efficiency barrier. In Bitcoin, Nakamoto consensus has the problem of being dependent on transaction/block synchronization. This motivates many proposals for parallel block generation with synchronization on public blockchains. One idea is to elect a leader or committee before generating blocks, such as Bitcoin-NG [45], Delegated Proof of Stake (DPOS) [61], Ouroboros [62], and Algorand [63]. Another option is to recognize miners' work on non-longest chains, such as Greedy Heaviest-Observed Sub-Tree (GHOST) protocol [64], Conflux [65], and IOTA [66].

On the other hand, consortium blockchains typically employ deterministic consensus algorithms, most of which are derived from BFT algorithms. The validity of blocks relies on message exchange rather than nodes' proof of work so that the generation time is reduced. Although the BFT algorithm provides good performance for small-scale applications, it does not suit large-scale applications due to its high communication cost and poor scalability. Speculative BFT [67] is proposed to reduce the communication complexity to $O(N)$ in the ideal case. However, it requires higher communication complexity when replacing the leader, making it inapplicable to practical blockchain networks. Further, Proof of Vote (PoV) [68,69] and Hotstuff [54] are proposed with linear consensus decision and view change.

Actually, all peers need to validate each transaction in a decentralized system. Therefore, $O(N)$ is the lower bound in terms of communication complexity. While leader or committee elections can break the guarantee bound in both public and consortium blockchains, the actual impact is different. In a public blockchain that is completely decentralized, in order to ensure fairness, a majority of nodes must participate in the consensus process. Therefore, the computing capacity and bandwidth requirements of the nodes should not be set too high, since this is not reliable. On the contrary, in a consortium blockchain, nodes are censored for access, therefore the governing agency can impose higher requirements on them. For example, core nodes must use commercial servers with strong computing and security capabilities, or they must follow a specific topology. In particular, for DNS alternative systems that provide identifier services, efficient and reliable consortium blockchains are more appropriate than public blockchains.

*(2) Challenge 2: Final Consistency*

The Nakamoto consensus represents a non-deterministic consensus in which nodes are allowed to generate successors of a block once they have solved the puzzle, which will result in forks. For the purpose of selecting the main chain, the longest chain rule is applied, which specifies that only one of the conflicting blocks with the same height is valid. However, as time passes, forks can turn strong consistency into final consistency.

Allowing forking involves balancing three properties in the CAP (Consistency, Availability, and Partition tolerance) trilemma [70]. Specifically, in the distributed theory, when the network topology is split into components by missing or failed links, a partition happens such that nodes in one component (partition) cannot communicate with nodes in another. Under this unfortunate circumstance, the distributed system designers are enforced to choose either availability or consistency. Public blockchains are characterized by consensus algorithms that focused on high availability and thus satisfy only final consistency requirements unlike consortium blockchains, which prioritize strong consistency. The BFT-based algorithms used by consortium blockchains reach consensus without forks. More than 2/3 of the core nodes commit the block deterministically in a consensus round through 2 or 3 phases of message exchange.

*(3) Challenge 3: Limits on Storage Capacity*

Nodes store a copy of all the on-chain data locally to independently process transactions and blocks. However, the exponential growth of blockchain poses a substantial challenge to the storage capacity of nodes. As an example, it would take 406.05 Gigabytes of disk space to copy all the Bitcoin data by July 10, 2022 [71]. In Bitcoin, light nodes are designed as a lightweight solution, which stores only block headers and verifies transactions by downloading blocks from full nodes. Due to the introduction of sharding, Ethereum further divides nodes into super-full, top-level, single-shard, and light nodes. In the absence of sufficient storage, nodes in Bitcoin or Ethereum can only become light nodes, affecting decentralization and security objectively.

The EC-based cooperative storage schemes [72,73], including the basic MIS, use erasure code to encode and decode data. Erasure code was originally a coding scheme for wireless channel transmission, and it was later used in distributed storage system to reduce duplications. The EC-based cooperative schemes reduce the on-chain data stored on nodes in a fixed proportion only by modifying the reading and writing processes of data incrementally. This makes such schemes suitable for both public and consortium blockchains.

The value of MIS in this paper is threefold. First, basic MIS provides a consortium blockchain solution that simultaneously manages multiple types of identifiers and is compatible with legacy DNS. As we move into the Web 3.0 era, MIS is well suited to interconnect diverse sub-metaverses. Second, due to the lack of sufficient decentralization on the consortium blockchain, which may be necessary for some scenarios, we propose an Ethereum-based modification called EMIS. The EMIS system abstracts and implements the identifier logic for management and resolution through Solidity smart contracts. Third, basic MIS utilizing consortium blockchain technology is capable of addressing the challenges mentioned above. In addition to its low latency and strong consistency, the PPoV consensus scheme is lightweight and reduces storage consumption. Table 3 below presents a comparative summary between MIS and different existing DNS alternative systems.

Table 3: Comparison between MIS and existing DNS alternatives.

| DNS Alternatives | Type of Blockchain | Consensus | Consistency | Multi-identifier Management | Compatibility of Legacy DNS | Average Consensus Latency (seconds) |
|---|---|---|---|---|---|---|
| Namecoin | Public | Nakamoto | Final | No | No | 2400 |
| Blockstack | Public | Nakamoto | Final | No | No | 600 |
| ENS | Public | PoS | Final | No | Yes | 13.38 |
| DNSTSM | Consortium | Kafka | Final | No | Yes | 0.095 |
| TD-Root | Consortium | C-RAND | Strong | No | Yes | <350 |

| MIS | Consortium | PPoV | Strong | Yes | Yes | 0.357 |
| EMIS | Public | PoS | Final | Yes | Yes | 13.38 |

# 8 Conclusion

In this paper, we propose MIS, the first multi-identifier management and resolution system for metaverse. As a DNS alternative, MIS eliminates the centralization of legacy DNS in the TCP/IP architecture. In order to separate on-chain and off-chain data, the basic MIS is designed as a 4-tier architecture on the consortium blockchain, under the assumption that the future metaverse will continue to evolve persistently using a wide range of sub-metaverses and different types of resources. Resource data is stored heterogeneously off-chain with a multi-authority attribute-based encryption scheme for privacy protection and access control. Several innovative improvements are also made to the size of data to reduce the storage pressure on consortium blockchain nodes. For decentralization prioritization, we implement the multi-identifier logic as EMIS on Ethereum. We have deployed two testbeds in 26 countries from 5 continents worldwide. Our experimental results show that consortium blockchain-based MIS outperforms legacy DNS and alternatives in terms of storage and latency. Finally, it should be mentioned that both the basic MIS and EMIS have been released online as open-source management systems for the MIN architecture.

### ACKNOWLEDGEMENTS

This work was supported by the National Keystone Research and Development Program of China [2017YFB0803204]; Foshan Innovation Team [2018IT100082]; Basic Research Enhancement Program of China [2021-JCJQ-JJ-0483]; China Environment for Network Innovation GJFGW [2020]386, SZFGW [2019]261; Guangdong Province Research and Development Key Program [2019B010137001]; Guangdong Province Basic Research [2022A1515010836]; Shenzhen Research Programs [JCYJ20220531093206015, JCYJ20210324122013036, JCYJ20190808155607340]; Shenzhen Fundamental Research Program [GXWD20201231165807007-20200807164903001]; ZTE Funding [2019ZTE03-01]; Huawei Funding [TC20201222002].

### REFERENCES

[1] Lik-Hang Lee, Tristan Braud, Pengyuan Zhou, Lin Wang, Dianlei Xu, Zijun Lin, Abhishek Kumar, Carlos Bermejo and Pan Hui. All one needs to know about metaverse: A complete survey on technological singularity, virtual ecosystem, and research agenda. arXiv e-prints (October 01 2021).

[2] Joe Sanchez. Second life: An interactive qualitative analysis. In Proceedings of the Society for Information Technology & Teacher Education International Conference 2007 (San Antonio, Texas, USA, 2007).

[3] John David N. Dionisio, William G. Burns III and Richard Gilbert. 3d virtual worlds and the metaverse: Current status and future possibilities. ACM Comput. Surv., 45, 3 (2013), 34.

[4] Anders Bruun and Martin Lynge Stentoft. Lifelogging in the wild: Participant experiences of using lifelogging as a research tool. In Proceedings of the Human-Computer Interaction – INTERACT 2019: 17th IFIP TC 13 International Conference (Cham, 2019).

[5] Kyoungro Yoon, Sang-Kyun Kim, Sangkwon Peter Jeong and Jeong-Hwan Choi. Interfacing cyber and physical worlds: Introduction to ieee 2888 standards. In Proceedings of the 2021 IEEE International Conference on Intelligent Reality (ICIR) (12-13 May 2021, 2021).

[6] Satoshi Nakamoto. Bitcoin: A peer-to-peer electronic cash system. Decentralized Business Review (2008), 21260.

[7] Ethereum, https://ethereum.org/en/.

[8] Namecoin, https://www.namecoin.org/.

[9] Aaron Swartz. Squaring the triangle: Secure, decentralized, human-readable names, http://www.aaronsw.com/weblog/squarezooko.

[10] Muneeb Ali, Jude Nelson, Ryan Shea and Michael J. Freedman. Blockstack: A global naming and storage system secured by blockchains. In Proceedings of the the 2016 USENIX Conference on Usenix Annual Technical Conference (Denver, CO, USA, 2016).

[11] Ens documentation. 2022, https://docs.ens.domains/.

[12] Greg Slepak. Dnschain+okturtles. 2014, https://okturtles.org/other/dnschain_okturtles_overview.pdf.

[13] Emercoin, https://emercoin.com/en/.

[14] Ittay Eyal and Emin Gün Sirer. Majority is not enough: Bitcoin mining is vulnerable. Communications of the ACM, 61, 7 (2014), 436-454.

[15] Zhong Yu, Dong Xue, Jiulun Fan and Chang Guo. Dnstsm: Dns cache resources trusted sharing model based on consortium blockchain. IEEE

Access, 8 (2020), 13640-13650.

[16] Guobiao He, Wei Su, Shuai Gao and Jiarui Yue. Td-root: A trustworthy decentralized dns root management architecture based on permissioned blockchain. Future Generation Computer Systems, 102 (2020), 912-924.

[17] Ho-Kyung Yang, Hyun-Jong Cha and You-Jin Song. Secure identifier management based on blockchain technology in ndn environment. IEEE Access, 7 (2018), 6262-6268.

[18] Hui Li, Jiangxing Wu, Xin Yang, Han Wang, Julong Lan, Ke Xu, Hua Tan, Jinwu Wei, Wei Liang and Fusheng Zhu. Min: Co-governing multi-identifier network architecture and its prototype on operator's network. IEEE Access, 8 (2020), 36569-36581.

[19] Welcome to decentraland, https://decentraland.org/.

[20] Cryptovoxels - a user owned virtual world, https://cryptowexels.com/.

[21] The sandbox game — user-generated crypto, https://www.sandbox.game.

[22] Haihan Duan, Jiaye Li, Sizheng Fan, Zhonghao Lin, Xiao Wu and Wei Cai. Metaverse for social good: A university campus prototype. In Proceedings of the MM '21: Proceedings of the 29th ACM International Conference on Multimedia (New York, NY, USA, October 2021, 2021).

[23] Aleksandar Jovanović and Aleksandar Milosavljević. Vortex metaverse platform for gamified collaborative learning. Electronics, 11, 3 (2022), 317.

[24] Zijian Bao, Wenbo Shi, Debiao He and Kim-Kwang Raymond Chood. Iotchain: A three-tier blockchain-based iot security architecture. arXiv e-prints (June 01, 2018 2018).

[25] Mohammed Amine Bouras, Qinghua Lu, Sahraoui Dhelim and Huansheng Ning. A lightweight blockchain-based iot identity management approach. Future Internet, 13, 2 (2021), 24.

[26] Youngjun Song and Sunghyuck Hong. Build a secure smart city by using blockchain and digital twin. International Journal of Advanced Science and Convergence, 3 (2021), 9-13.

[27] Chunpeng Ge, Zhe Liu, Jinyue Xia and Liming Fang. Revocable identity-based broadcast proxy re-encryption for data sharing in clouds. IEEE Transactions on Dependable and Secure Computing, 18, 3 (2019), 1214-1226.

[28] Chunpeng Ge, Willy Susilo, Joonsang Baek, Zhe Liu, Jinyue Xia and Liming Fang. A verifiable and fair attribute-based proxy re-encryption scheme for data sharing in clouds. IEEE Transactions on Dependable and Secure Computing, 19, 5 (2021), 2907-2919.

[29] Matt Blaze, Gerrit Bleumer and Martin Strauss. Divertible protocols and atomic proxy cryptography. In Proceedings of the Advances in Cryptology—EUROCRYPT'98: International Conference on the Theory and Application of Cryptographic Techniques Espoo, Finland, May 31–June 4, 1998 Proceedings 17 (1998).

[30] Adi Shamir. Identity-based cryptosystems and signature schemes. In Proceedings of the Advances in Cryptology: Proceedings of CRYPTO 84 4 (1985).

[31] Amit Sahai and Brent Waters. Fuzzy identity-based encryption. In Proceedings of the Advances in Cryptology–EUROCRYPT 2005: 24th Annual International Conference on the Theory and Applications of Cryptographic Techniques, Aarhus, Denmark, May 22-26, 2005. Proceedings 24 (2005).

[32] John Bethencourt, Amit Sahai and Brent Waters. Ciphertext-policy attribute-based encryption. In Proceedings of the 2007 IEEE symposium on security and privacy (SP'07) (2007).

[33] Chunpeng Ge, Willy Susilo, Zhe Liu, Jinyue Xia, Pawel Szalachowski and Liming Fang. Secure keyword search and data sharing mechanism for cloud computing. IEEE Transactions on Dependable and Secure Computing, 18, 6 (2020), 2787-2800.

[34] Chunpeng Ge, Willy Susilo, Joonsang Baek, Zhe Liu, Jinyue Xia and Liming Fang. Revocable attribute-based encryption with data integrity in clouds. IEEE Transactions on Dependable and Secure Computing, 19, 5 (2021), 2864-2872.

[35] Xiaohui Liang, Rongxing Lu, Xiaodong Lin and Xuemin Sherman Shen. Ciphertext policy attribute based encryption with efficient revocation. 2010, http://bbcr.uwaterloo.ca/~x27liang/abe_with_revocation.pdf.

[36] Hui Cui and Robert H. Deng. Revocable and decentralized attribute-based encryption. The Computer Journal, 59, 8 (2016), 1220-1235.

[37] Jiawei Zhang, Jianfeng Ma, Yanbo Yang, Ximeng Liu and Neal N. Xiong. Revocable and privacy-preserving decentralized data sharing framework for fog-assisted internet of things. IEEE Internet of Things Journal, 9, 13 (2021), 10446-10463.

[38] Ping Yu, Qiaoyan Wen, Wei Ni, Wenmin Li, Caijun Sun, Hua Zhang and Zhengping Jin. Decentralized, revocable and verifiable attribute-based encryption in hybrid cloud system. Wireless Personal Communications, 106 (2019), 719-738.

[39] Allison Lewko and Brent Waters. Decentralizing attribute-based encryption. In Proceedings of the Advances in Cryptology–EUROCRYPT 2011: 30th Annual International Conference on the Theory and Applications of Cryptographic Techniques (Tallinn, Estonia, May 15-19, 2011).

[40] Ipfs powers the distributed web, https://ipfs.io.

[41] Michael Dowling. Is non-fungible token pricing driven by cryptocurrencies? Finance Research Letters, 44 (2022/01/01/ 2022), 102097.

[42] Catalina Goanta. Selling land in decentraland: The regime of non-fungible tokens on the ethereum blockchain under the digital content directive. Springer International Publishing, 2020.

[43] Brendan Benshoof, Andrew Rosen, Anu G. Bourgeois and Robert W. Harrison. Distributed decentralized domain name service. In Proceedings of the 2016 IEEE International Parallel and Distributed Processing Symposium Workshops (IPDPSW) (2016).

[44] Handshake, https://hsd-dev.org/.

[45] Ittay Eyal, Adem Efe Gencer, Emin Gün Sirer and Robbert Van Renesse. Bitcoin-ng: A scalable blockchain protocol. In Proceedings of the the 13th Usenix Conference on Networked Systems Design and Implementation (Santa Clara, CA, 2016).

[46] Muneeb Ali. Stacks 2.0: Apps and smart contracts for bitcoin. 2020, https://gaia.blockstack.org/hub/1Eo6q4qLMcSSpkhoUADxRAGZhgUyjVEVcK/stacks-zh.pdf.

[47] Wentong Wang, Ning Hu and Xin Liu. Blockzone: A blockchain-based dns storage and retrieval scheme. In Proceedings of the International Conference on Artificial Intelligence and Security (Cham, 2019).

[48] Miguel Castro and Barbara Liskov. Practical byzantine fault tolerance. In Proceedings of the OSDI '99: Proceedings of the third symposium on Operating systems design and implementation (1999).

[49] Tong Jin, Xiang Zhang, Yirui Liu and Kai Lei. Blockndn: A bitcoin blockchain decentralized system over named data networking. In Proceedings of the 2017 Ninth international conference on ubiquitous and future networks (ICUFN) (2017).

[50] Jingqiang Liu, Bin Li, Lizhang Chen, Meng Hou, Feiran Xiang and Peijun Wang. A data storage method based on blockchain for decentralization dns. In Proceedings of the 2018 IEEE Third International Conference on Data Science in Cyberspace (DSC) (18-21 June 2018, 2018).

[51] Wondeuk Yoon, Indal Choi and Daeyoung Kim. Blockons: Blockchain based object name service. In Proceedings of the 2019 IEEE International Conference on Blockchain and Cryptocurrency (ICBC) (2019).

[52] Yantao Shen, Yang Lu, Zhili Wang, Xin Xv, Feng Qi, Ningzhe Xing and Ziyu Zhao. Dns service model based on permissioned blockchain. Intelligent Automation and Soft Computing, 27, 1 (2021), 259-268.

[53] Xiangui Wang, Kedan Li, Hui Li, Yinghui Li and Zhiwei Liang. Consortiumdns: A distributed domain name service based on consortium chain. In Proceedings of the 2017 IEEE 19th International Conference on High Performance Computing and Communications; IEEE 15th International Conference on Smart City; IEEE 3rd International Conference on Data Science and Systems (HPCC/SmartCity/DSS) (2017).

[54] Maofan Yin, Dahlia Malkhi, Michael K. Reiter, Guy Golan Gueta and Ittai Abraham. Hotstuff: Bft consensus with linearity and responsiveness. In Proceedings of the the 2019 ACM Symposium on Principles of Distributed Computing (Toronto ON, Canada, 2019).

[55] Yongjie Bai, Yang Zhi, Hui Li, Han Wang, Ping Lu and Chengtao Ma. On parallel mechanism of consortium blockchain: Take pov as an example. In Proceedings of the 2021 The 3rd International Conference on Blockchain Technology (Shanghai, China, 2021).

[56] Dan Boneh, Ben Lynn and Hovav Shacham. Short signatures from the weil pairing. In Proceedings of the International conference on the theory and application of cryptology and information security (2001).

[57] Stephen B. Wicker and Vijay K. Bhargava. Reed-solomon codes and their applications. John Wiley & Sons, Published, 1999.

[58] Mis-blockchain, https://github.com/MIN-Group/mis.

[59] Emis, https://github.com/MIN-Group/EMIS.

[60] https://www.speedtest.cn/.

[61] Delegated proof of stake (dpos). 2019, https://how.bitshares.works/en/master/technology/dpos.html.

[62] Aggelos Kiayias, Alexander Russell, Bernardo David and Roman Oliynykov. Ouroboros: A provably secure proof-of-stake blockchain protocol. In Proceedings of the Annual International Cryptology Conference (Cham, 2017).

[63] Yossi Gilad, Rotem Hemo, Silvio Micali, Georgios Vlachos and Nickolai Zeldovich. Algorand: Scaling byzantine agreements for cryptocurrencies. In Proceedings of the the 26th Symposium on Operating Systems Principles (Shanghai, China, 2017).

[64] Yonatan Sompolinsky and Aviv Zohar. Secure high-rate transaction processing in bitcoin. In Proceedings of the International Conference on Financial Cryptography and Data Security (Berlin, Heidelberg, 2015).

[65] Chenxin Li, Peilun Li, Dong Zhou, Zhe Yang, Ming Wu, Guang Yang, Wei Xu, Fan Long and Andrew Chi-Chih Yao. A decentralized blockchain with high throughput and fast confirmation. In Proceedings of the 2020 USENIX Annual Technical Conference (USENIX ATC 20) (2020).

[66] Serguei Popov. The tangle. 2018, https://assets.ctfassets.net/r1dr6vzfxhev/2t4uxvsIqk0EUau6g2sw0g/45eae33637ca92f85dd9f4a3a218e1ec/iota1_4_3.pdf.

[67] Ramakrishna Kotla, Lorenzo Alvisi, Mike Dahlin, Allen Clement and Edmund Wong. Zyzzyva: Speculative byzantine fault tolerance. In Proceedings of the twenty-first ACM SIGOPS symposium on Operating systems principles (Stevenson, Washington, USA, 2007).

[68] Kejiao Li, Hui Li, Hanxu Hou, Kedan Li and Yongle Chen. Proof of vote: A high-performance consensus protocol based on vote mechanism & consortium blockchain. IEEE, 2017.

[69] Kejiao Li, Hui Li, Han Wang, Huiyao An, Ping Lu, Peng Yi and Fusheng Zhu. Pov: An efficient voting-based consensus algorithm for consortium blockchains. Frontiers in Blockchain, 3, 11 (2020-March-18 2020).

[70] Seth Gilbert and Nancy Lynch. Brewer's conjecture and the feasibility of consistent, available, partition-tolerant web services. SIGACT News, 33, 2 (2002), 51–59.

[71] Size of the bitcoin blockchain from january 2009 to july 11, 2022(in gigabytes), https://www.statista.com/statistics/647523/worldwide-bitcoin-blockchain-size/.

[72] Doriane Perard, Jérôme Lacan, Yann Bachy and Jonathan Detchart. Erasure code-based low storage blockchain node. In Proceedings of the 2018 IEEE International Conference on Internet of Things (iThings) and IEEE Green Computing and Communications (GreenCom) and IEEE Cyber, Physical and Social Computing (CPSCom) and IEEE Smart Data (SmartData) (2018).

[73] Xiaodong Qi, Zhao Zhang, Cheqing Jin and Aoying Zhou. Bft-store: Storage partition for permissioned blockchain via erasure coding. In Proceedings of the 2020 IEEE 36th International Conference on Data Engineering (ICDE) (2020).